\documentclass[10pt,twocolumn]{article}
\usepackage[top=1.9cm, bottom=1.9cm, left=1.8cm, right=1.8cm]{geometry}

\usepackage{amsmath,amssymb,amsfonts}
\usepackage{listings}                       %
\usepackage[noend]{algpseudocode}
\usepackage{multirow}
\usepackage{graphicx}
\usepackage{textcomp}
\usepackage{xcolor}
\usepackage{flushend}
\usepackage[lined,boxed]{algorithm2e}

\usepackage{graphicx,subfigure}
\usepackage[bf]{caption}
\usepackage[square,sort,numbers]{natbib}

\SetCommentSty{mycommfont}

\usepackage{xcolor}
\usepackage{booktabs}
\usepackage{graphicx}
\usepackage{paralist}

\usepackage{times}

\usepackage[compact]{titlesec}
\titlespacing*{\section}{0pt}{*3}{3pt}
\titlespacing{\subsection}{0pt}{*2}{2pt}
\titlespacing{\subsubsection}{0pt}{*3}{3pt}

\definecolor{linkcol}{rgb}{0,0,0.5}
\definecolor{citecol}{rgb}{0,0.5,0.3}
\definecolor{urlcol}{rgb}{0.3,0,0}

\usepackage{xspace}

\let\OLDthebibliography\thebibliography
\renewcommand\thebibliography[1]{
  \OLDthebibliography{#1}
  \setlength{\parskip}{0pt}
  \setlength{\itemsep}{0pt plus 0.3ex}
}

\usepackage[hang,flushmargin]{footmisc}

\renewcommand{\footnoterule}{%
  \kern -3pt
  \hrule width 1in
  \kern 2pt
}

\usepackage{url}
\makeatletter
\def\url@leostyle{%
  \@ifundefined{selectfont}{\def\UrlFont{}}%
  {\def\UrlFont{}}%
}
\makeatother
\usepackage[hyphenbreaks]{breakurl}

\definecolor{darkred}{RGB}{153,0,0}
\definecolor{darkblue}{RGB}{0,0,99}
\usepackage[colorlinks=true, linkcolor = darkred,   citecolor = darkred, urlcolor = darkblue]{hyperref}

\newtheorem{ldp-definition}{Definition}
\newtheorem{dp-definition}{Definition}
\newcommand{\descr}[1]{\smallskip\noindent\textbf{#1}}

\title{\bf Non-Polar Opposites: Analyzing the Relationship Between Echo Chambers and Hostile Intergroup Interactions on Reddit\footnote{To appear at ICWSM 2023, please cite accordingly. Corresponding author: alexandros.efstratiou.20@ucl.ac.uk}}

\author {
    Alexandros Efstratiou,\textsuperscript{\rm 1}
    Jeremy Blackburn,\textsuperscript{\rm 2}
    Tristan Caulfield,\textsuperscript{\rm 1}\\
    Gianluca Stringhini,\textsuperscript{\rm 3}
    Savvas Zannettou,\textsuperscript{\rm 4}
    Emiliano De Cristofaro\textsuperscript{\rm 1}\\[1ex]
    \textsuperscript{\rm 1}University College London,
    \textsuperscript{\rm 2}Binghamton University,
    \textsuperscript{\rm 3}Boston University,
    \textsuperscript{\rm 4}TU Delft}

\date{}

\begin{document}

\maketitle

\begin{abstract}
Previous research has documented the existence of both online echo chambers and hostile intergroup interactions.
In this paper, we explore the relationship between these two phenomena by studying the activity of 5.97M Reddit users and 421M comments posted over 13 years.
We examine whether users who are more engaged in echo chambers are more hostile when they comment on other communities.
We then create a typology of relationships between political communities based on whether their users are toxic to each other, whether echo chamber-like engagement with these communities is associated with polarization, and on the communities' political leanings. 

We observe both the echo chamber and hostile intergroup interaction phenomena, but neither holds universally across communities.
Contrary to popular belief, we find that polarizing and toxic speech is more dominant between communities on the same, rather than opposing, sides of the political spectrum, especially on the left; however, this mainly points to the collective targeting of political outgroups.
\end{abstract}

\section{Introduction}\label{sec:intro}

In {\em echo chambers}, users encounter view-affirming information or other users, thus never experiencing any informational disruption~\cite{sunstein_republiccom_2001}.
Users tend to engage with communities that politically align with their views~\cite{waller_quantifying_2021}, while controversial events often lead to spontaneously formed polarized networks~\cite{barbera_tweeting_2015,del_vicario_public_2017}.
At the same time, users who try to bridge opposing views tend to receive lower attention and social rewards~\cite{garimella_political_2018}.
Diversifying user exposure and diminishing such echo chamber spaces is a promising approach to securing the integrity of deliberative democracy~\cite{matakos_maximizing_2020}.

{\em Hostile intergroup interactions} pose a challenge to this approach.
Participation in mixed social media networks~\cite{vaccari_echo_2016}, diverse media diets~\cite{guess_why_2018}, and encountering disagreeable views~\cite{dubois_echo_2018} are all fairly common.
When interactions between users on opposing camps do occur, however, they tend to be more toxic and hostile~\cite{de_francisci_morales_no_2021,cinelli_echo_2021,bliuc_you_2020,marchal_be_2021}.

\descr{Problem statement.} We hypothesize that echo chambers and hostile interactions may not be mutually exclusive.
Instead, one may influence the degree to which the other occurs.
Nonetheless, this has yet to be explored.

Moreover, users may display varying degrees of engagement with their ``echo chambers.''
Research that analyzes echo chambers at the community level may thus not capture this.
Here, we set out to recognize such understudied differences in engagement using a user-level approach. %

Overall, we focus on two main research questions:
\begin{enumerate}
  \item[\textbf{RQ1}] %
  How is a user's degree of engagement with a political echo chamber related to their hostility in an intergroup interaction?
  \item[\textbf{RQ2}] What are the different relationships between Reddit political communities based on the hostility and the polarization of their user base, and how do they vary depending on their political leanings?
\end{enumerate}

\descr{Methodology.}
Our work builds on a dataset of 421M comments made between 2006 and 2019 from 5.97M unique authors on Reddit.
These appeared across 918 political subreddits, which we cluster into distinct ``echo chamber'' communities based on their user similarities (see Section~\ref{sec:communities}).
We allocate users into home communities by analyzing where they were most active over these 13 years %
and measure the toxicity of the comments left by each community's home users on the other communities (Section~\ref{sec:main_analysis}).

For this study, any community can be an echo chamber for a given user if they only (or disproportionately) engage with it. 
Therefore, we define echo chamber engagement as the proportion of comments that a user left in their preferred community.
This definition approximates a user's preference for homophily~\cite{rogers_homophily-heterophily_1970}, i.e., their tendency to surround themselves with similar others.
We define an intergroup interaction as the event of a user leaving their echo chamber to comment on another community.
We treat the interaction as hostile if the comment is toxic (as determined through Google's Perspective API).

For {\bf RQ1}, we use mixed-effects logistic models to assess how the probability that a user's comment will be toxic on some target community is related to the proportion of comments left by the user in their echo chamber.
We then combine our cross-toxicity and mixed-effects analyses to create a typology of community relationships (Section~\ref{sec:typology}) and observe the frequency of each type based on a manual assessment of the communities' political leanings (i.e., whether they are on the same or opposing sides) to address {\bf RQ2}.

\descr{Main findings.}
Overall, Reddit's political space between 2006-2019 included 16 communities not captured by a binary left-right split.
Users mainly engaged with their home communities in line with the echo chamber narrative; however, they also posted a non-negligible proportion of comments on other communities overall.
Some of these communities were almost universally toxic (or non-toxic), while others showed more selectivity in {\em where} they were toxic.
We also show that increased engagement with these ``echo chamber'' communities had differential relationships to hostility outside of them, depending on the target community.
Toxic behavior was up to 2.5 times more likely with higher echo chamber engagement when the relationship between communities was polarizing and down to nearly 70 times less likely when the relationship was depolarizing; however, this depolarization may also be attributable to content moderation.

We do not find universal tribalism on Reddit.
Specifically, the most common type of relationship (21\%) was an indifferent one.
Contrary to conventional wisdom~\cite{de_francisci_morales_no_2021,marchal_be_2021}, inciting and polarizing relationships were more common between communities on the same (6\%) rather than on opposing sides (2\%) of the political spectrum.

Our contributions are threefold.
First, we begin to {\em systematically} typologize community relationships.
This provides a more accurate map of the state of political discourse, including understudied elements such as indifferent communities, polarizing relationships between communities of similar leanings, and civil relationships between communities of opposite leanings.
Second, we reveal that whereas increased engagement with some communities is indeed associated with increased hostility toward others, the opposite relationship holds for several communities.
This can allow future research to better target anti-polarization interventions like diversified exposure.
For example, increasing network diversity should be a promising approach in cases where higher echo chamber engagement is related to higher hostility but may fail in cases where the opposite is true.
Finally, we open future research directions for the role of moderation in these observations.

\section{Background and Related Work}

In this section, we cover work on echo chambers, hostile intergroup interactions, and gaps in attempts to link the two.\smallskip

\subsection{Echo Chambers}
Echo chambers are relatively widespread on social media~\cite{terren_echo_2021}.
In terms of the content that users are exposed to, roughly 90\% of the political videos that the average user consumes on YouTube align with their political beliefs~\cite{hosseinmardi_evaluating_2020}.
Furthermore, science-advocating Facebook users tend to only interact with scientific pages, whereas conspiratorial users only interact with conspiratorial pages; users interacting with both kinds of pages are very rare~\cite{brugnoli_recursive_2019}.
While several fact-checks are aimed toward these conspiratorial users,~\citet{zollo_debunking_2017} find that only about 1.2\% of them interact with this information.

Echo chambers may also arise via interactions with similar users.
\citet{garimella_political_2018} find high polarity in political networks on Twitter, with highly partisan users receiving more engagement.
On Reddit, users tend to interact with ideologically similar communities, aka {\em subreddits}~\cite{waller_quantifying_2021}.
However,~\citet{de_francisci_morales_no_2021}, looking at {\em r/politics}, which is one of the largest political subreddits during the 2016 election, find that cross-cutting interactions are pretty common there.

Specific platform affordances may play a role in the formation of echo chambers.
In a platform comparison study,~\citet{cinelli_echo_2021} find that echo chambers are more prominent on Facebook and Twitter than Reddit.
This may be because Facebook and Twitter use recommender algorithms more frequently, resulting in so-called {\em filter bubbles}~\cite{pariser_filter_2011}.
Indeed,~\citet{bakshy_exposure_2015} find that introducing algorithmic ranking of content can reduce the exposure of Facebook users to cross-cutting content, although individual user choice has a more significant effect on this.
On Spotify, recommendations can reduce the overall diversity of podcasts that individual users engage with~\cite{holtz_engagement-diversity_2020}.
A simulation study finds that several different types of recommender algorithms %
can increase the similarity of content that already similar users engage with~\cite{chaney_how_2018}.

Echo chambers may be exacerbated by controversy around a given topic.
Controversial events which cause nationwide debates eventually lead to echo chamber discussions between users of similar beliefs~\cite{barbera_tweeting_2015,del_vicario_public_2017}.
Radicalization through similar content exposure is another factor; for example,~\citet{hosseinmardi_evaluating_2020} report a surge in alt-right video consumption on YouTube, with radicalization occurring only for right-wing users.
Similarly,~\citet{ribeiro_auditing_2020} show that initial consumption of mild right-leaning content can lead to eventual consumption of far-right content.

Overall, echo chambers may alienate users to specific points of view, making them apprehensive of such opinions when encountering them.
The small fraction of conspiratorial users who interact with fact-checks in~\citet{zollo_debunking_2017} become more polarized following this exposure.
Relatedly, the higher users' activity in their preferred spaces, the more polarized these users become~\cite{brugnoli_recursive_2019}.
Therefore, disproportional interaction with only specific kinds of content or users may affect behavior upon interaction with other kinds.

\subsection{Intergroup interactions}
When partisans witness criticism against their side, they wish to distance themselves from opposing partisans~\cite{suhay_polarizing_2018}.
However, they also overestimate how much the latter is prejudiced against them~\cite{moore-berg_exaggerated_2020}, and correcting these perceptions can reduce political intergroup prejudice~\cite{lees_inaccurate_2020}.
Thus, engaging with oppositional users can both increase polarization (if the user witnesses criticism) or decrease it (if perceptions of prejudice are corrected); research so far mainly supports the former.

Two separate studies on the r/politics community on Reddit find that cross-partisan interactions are pretty common but tend to be more hostile~\cite{de_francisci_morales_no_2021,marchal_be_2021}.
A YouTube case study of a controversial video finds that users in the comment section often engage in hostile interactions with users of opposing views~\cite{bliuc_you_2020}.

Such interactions are not always naturally occurring.
Some Reddit users have ``anti-social homes'' where they go to display elevated hostility~\cite{datta_extracting_2019}.
Moreover,~\citet{kumar_community_2018} find that users on certain subreddits initiate negative mobilizations on others by posting links targeting posts in other communities. %
``Brigading attacks'', i.e., targeting another community to down-vote posts and harass its users, also occur on Reddit~\cite{mills_pop-up_2018}.

Hostile intergroup interactions may be elevated during election periods~\cite{datta_extracting_2019}, and toxicity is higher when political discussions occur in explicitly political rather than non-political Reddit spaces~\cite{rajadesingan_political_2021}.
When such hostile interactions occur strictly across political divides, this may be because the views or opinions of a disagreeable user can be construed as threats toward a counter-partisan's political social identity~\cite{van_bavel_partisan_2018}, which may introduce motivations to protect this identity (often in the form of hostility toward the outgroup).
The norms of a given community also play a role in the prominence of toxic content there~\cite{rajadesingan_quick_2020}.

Interacting with outgroups can have various effects.
Twitter users who are asked to follow bots posting oppositional content become even more entrenched in their prior views 1.5 months later~\cite{bail_exposure_2018}.
Similarly, fact-checks aimed toward conspiratorial Facebook users seem to backfire and drive more conspiratorial content engagement~\cite{zollo_debunking_2017}.
On Reddit, negative interactions with outgroup members reduce the likelihood that such cross-cutting interactions will reoccur in the future~\cite{marchal_be_2021}.
On the other hand, sports fans who engage in cross-cutting interactions use more problematic language in their teams' communities~\cite{zhang_intergroup_2019}.

\subsection{Remarks} 
Although several studies have outlined how increased echo chamber engagement may drive higher polarization~\cite{brugnoli_recursive_2019} and radicalization~\cite{hosseinmardi_evaluating_2020,ribeiro_auditing_2020}, as well as how cross-cutting exposure may drive higher preference for echo chamber-like consumption~\cite{bail_exposure_2018,barbera_tweeting_2015,del_vicario_public_2017}, it remains unclear whether the degree of echo chamber engagement is related to the subsequent hostility expressed in intergroup interactions.
To our knowledge, we are the first to study this.
We examine the bulk of Reddit's political sphere to understand the dynamics between engagement, hostility, and the political leanings of different communities.

\section{Dataset and Political Communities}\label{sec:communities}

In this section, we present our dataset and how we cluster subreddits to identify distinct political communities.

\subsection{Data Sources} 
Our starting point is a list of 31K subreddits, curated by~\citet{rajadesingan_political_2021}, labeled based on the percentage of political comments they host.
We treat political subreddits as those hosting 50\% or more political content and retain those with at least 1,000 comments and made by at least 100 unique authors.
This leaves 918 subreddits.

\begin{figure}[t!]
    \centering
    \includegraphics[width=0.9\columnwidth]{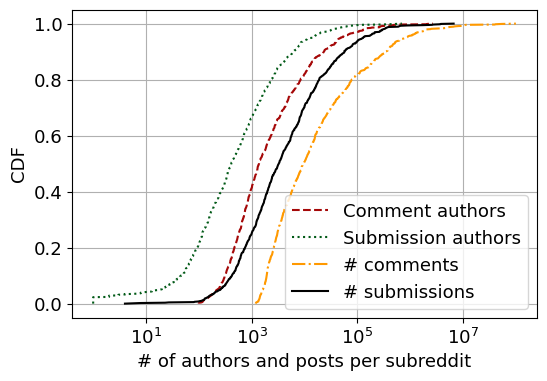}
    \caption{Cumulative Distribution Function (CDF) of the numbers of unique authors and posts for submissions and comments per subreddit.}
    \label{fig:subs_size}
    \vspace{-0.3cm}
\end{figure}

We obtain all comments posted in these 918 subreddits between June 12th, 2006, and December 31st, 2019, using the Pushshift Reddit dataset~\cite{baumgartner_pushshift_2020}.
Thus, our analyses are historical and may not reflect the current state of Reddit.
Overall, we examine 421M comments from 5.97M authors.
In Figure~\ref{fig:subs_size}, we plot the Cumulative Distribution Function (CDF) of the number of comments, comment authors, submissions, and submission authors.
The normal distributions suggest that our subreddits provide an adequate approximation of Reddit's political sphere across spaces with varying degrees of engagement.

\descr{Ethical considerations.}
This project received ethical approval from UCL's Research Ethics Committee (Project ID: 19379/001).
Note that we do not attempt to identify any users appearing in our dataset beyond the use of unique pseudonyms (usernames) to identify comments made by the same user.
We only collect and analyze the minimum required amount of data for our research questions.

\begin{figure}[t!]
    \centering
    \includegraphics[width=0.95\columnwidth]{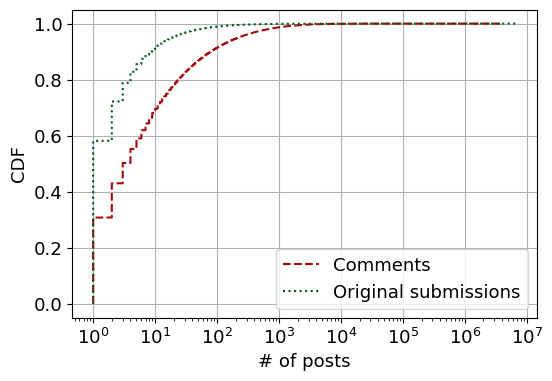}
    \caption{Cumulative Distribution Function (CDF) of comments and submissions across all subreddits per author.}
    \label{fig:author_cdf}
    \vspace{-0.2cm}
\end{figure}

\begin{table*}[t!]
  \centering
  \small
  \begin{tabular}{lrlrrr}
  \toprule
      \textbf{Community (abbreviation)} & \textbf{Size} & \textbf{Indicative subreddits} & \textbf{\#Comments} & \textbf{\#Users} & \textbf{\#Home users}\\
      \midrule
      Center-left (CL) & 144 & r/politics, r/Liberal, r/obama & 236,568,074 & 4,785,269 & 2,898,962\\
      Pro-Trump (TR) & 68 & r/The\_Donald, r/Infowars & 47,776,673 & 735,728 & 256,914\\
      EU/UK Politics (EU-UK) & 38 & r/unitedkingdom, r/europeans & 33,353,351 & 707,332 & 243,473\\
      Socialist (SOC) & 89 & r/MurderedByAOC, r/SandersForPresident & 17,429,739 & 575,307 & 105,803\\
      Middle East/World conflicts (ME) & 71 & r/Israel, r/antiwar & 15,994,336 & 683,431 & 137,894\\
      Anti-Trump (NoTR) & 85 & r/The\_Mueller, r/MarchAgainstTrump & 13,802,750 & 910,826 & 132,073\\
      Libertarian (LIB) & 50 & r/Libertarian, r/ronpaul & 13,553,498 & 483,583 & 76,577\\
      Pro-Democrat (DEM) & 34 & r/hillaryclinton, r/JoeBiden & 9,375,513 & 184,014 & 21,390\\
      Left-wing (LEFT) & 63 & r/communism, r/BlackLivesMatter & 7,740,333 & 445,084 & 77,293\\
      Anti-political extremes (NoEX) & 21 & r/stupidpol, r/InternetHitlers & 6,584,089 & 312,920 & 27,443\\
      Conservative (CON) & 27 & r/Republican, r/Conservative & 5,623,669 & 251,153 & 26,619\\
      Intellectual Dark Web (IDW) & 63 & r/JordanPeterson, r/daverubin & 5,563,082 & 321,540 & 69,885\\
      Alt-right (ALTR) & 48 & r/new\_right, r/WhiteNationalism & 2,835,060 & 174,035 & 23,438\\
      Gun discussions (GUN) & 25 & r/GunsAreCool, r/GunResearch & 2,524,063 & 134,913 & 24,439\\
      Automatic News (AUTO) & 45 & r/GUARDIANauto, r/Fox\_Nation & 1,278,417 & 43,575 & 3,773\\
      Model politics (MOD) & 40 & r/ModelUSGov, r/MHOC & 920,556 & 18,523 & 4,454\\
      \bottomrule
  \end{tabular}
  \caption{List of communities after community detection. Size refers to \# subreddits clustered in the respective community.}
  \label{tab:comm_labels}
  \vspace{-0.3cm}
\end{table*}

\subsection{Author Similarity Computation}

We cluster individual subreddits into larger communities based on their author similarities.
If communities share the same users, they might also host the same kinds of opinions, forming \textit{potential} ``echo chambers'' for this study.

To obtain author similarities between subreddits, we follow a similar approach to~\citet{datta_identifying_2017}.
First, we create Term Frequency-Inverse Document Frequency (TF-IDF) bag-of-word vectors, where each term is a unique author and each document is an individual subreddit.
We then filter out authors with a TF (number of comments) of less than 10 (we choose 10 informed by Figure~\ref{fig:author_cdf}, which shows that approximately 70\% of authors post fewer than ten comments).
We also filter document frequency to keep authors who have posted to at least 4 (i.e., median value) and no more than 2.5\% (22) of the total subreddits in our dataset. 
These filters prevent highly sparse vectors and retain only {\em active} authors whose commenting diversity is informative.
The TF-IDF vocabulary, therefore, includes all non-filtered authors.
Finally, we compute pairwise cosine similarities between the subreddit TF-IDF vectors.

\begin{figure*}[!t]
    \centering
    \includegraphics[width=0.99\textwidth]{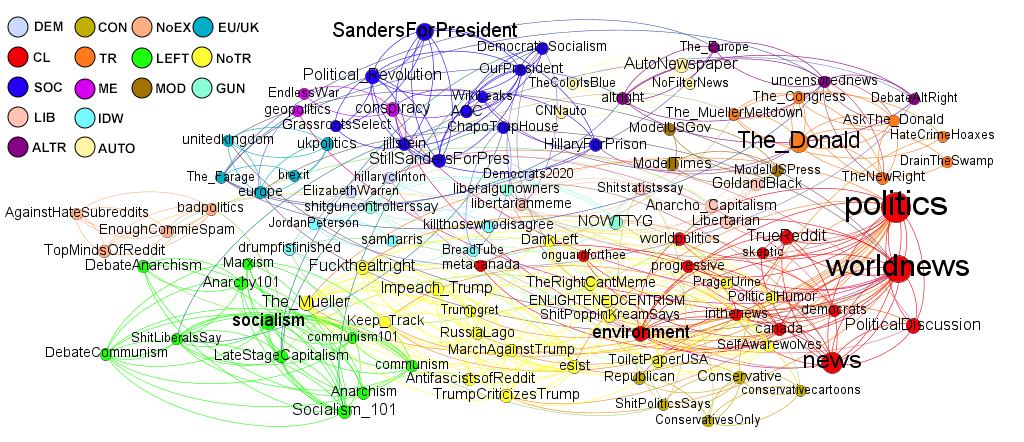}
    \caption{Similar subreddit network with top 10\% nodes in terms of degree. Nodes belong to the corresponding community in the legend.}
    \label{fig:network}
        \vspace{-0.3cm}
\end{figure*}

\subsection{Community Detection}

Next, {as per~\citet{datta_identifying_2017}, we build a subreddit network using the top 1\% cosine similarity values per subreddit as retained undirected edges.
We drop the bottom 5\% of these edges across \textit{all} subreddits to filter out arbitrary connections~\cite{datta_identifying_2017,von_luxburg_tutorial_2007}.
Finally, we apply the Louvain algorithm~\cite{blondel_fast_2008} to detect subreddit communities.
Louvain maximizes the density {\em within} and minimizes the density {\em between} communities~\cite{newman_modularity_2006}.
We obtain a modularity value of 0.58 using this approach, indicating good clustering~\cite{clauset_finding_2004}.

This yields 16 distinct communities; see Table~\ref{tab:comm_labels}.
Some of these are on the same side of the political spectrum but hold differing viewpoints or have different foci (e.g., distinct pro-Democrat and pro-Socialist communities, distinct pro-Conservative and pro-Trump communities, etc.).
From a manual inspection of the communities, we label six communities as {\em left-leaning} (center-left, pro-Democrat, left-wing, anti-extremism, Socialist, anti-Trump), five as {\em neutral} (EU/UK politics, Middle East/world conflicts, autonews, guns, model politics), and five as {\em right-leaning} (Intellectual Dark Web, pro-Trump, Conservative, alt-right, Libertarian).
Figure~\ref{fig:network} illustrates the subreddit network retaining nodes with degrees in the top 10\%.
Some communities, e.g., Socialist (SOC), are predominantly made up of advocation subreddits that support specific candidates (e.g., SandersForPresident).
Others, e.g., pro-Trump (TR) and anti-Trump (NoTR), are adversarial, where some subreddits are explicit ``responses'' to others (e.g., The\_Mueller in NoTR vs. The\_MuellerMeltdown in TR).
In Table~\ref{tab:network_stats}, we provide basic statistics for the overall network and per community.
We also provide the complete list of the 918 subreddits along with the communities they are allocated to in a Google document.\footnote{Please see \href{https://docs.google.com/document/d/1XVuHP96zcnrcMqOfEtD3oJ9vmsy8DDG9x8qNYtkz9SU}{https://docs.google.com/document/d/\\1XVuHP96zcnrcMqOfEtD3oJ9vmsy8DDG9x8qNYtkz9SU}}
Note that some subreddits have been banned or restricted, and 24 were banned during our observation period.
Banned or restricted subreddits are highlighted in the Google document.

\begin{table}[t!]
    \centering
    \small
    \begin{tabular}{lrrrrr}
    \toprule
        \textbf{Name} & \textbf{Av. Deg.} & \textbf{$D$} & \textbf{Dens.} & \textbf{Av. $C$} & \textbf{Av. PL}\\
        \midrule
        CL & 8.26 & 6 & 0.058 & 0.35 & 2.64 \\
        TR & 6.88 & 5 & 0.103 & 0.42 & 2.40 \\
        EU-UK & 8.74 & 4 & 0.236 & 0.60 & 1.99 \\
        SOC & 8.63 & 5 & 0.098 & 0.44 & 2.48 \\
        ME & 6.17 & 6 & 0.088 & 0.37 & 2.99 \\
        NoTR & 8.38 & 6 & 0.100 & 0.41 & 2.56 \\
        LIB & 7.00 & 6 & 0.143 & 0.58 & 2.50 \\
        DEM & 5.94 & 5 & 0.180 & 0.57 & 2.49 \\
        LEFT & 8.76 & 5 & 0.141 & 0.54 & 2.26 \\
        NoEX & 9.43 & 3 & 0.471 & 0.79 & 1.65 \\
        CON & 6.30 & 5 & 0.242 & 0.65 & 2.19 \\
        IDW & 5.97 & 6 & 0.096 & 0.40 & 2.91 \\
        ALTR & 6.50 & 7 & 0.138 & 0.49 & 2.73 \\
        GUN & 7.44 & 4 & 0.310 & 0.71 & 2.09 \\
        AUTO & 7.87 & 4 & 0.179 & 0.54 & 2.16 \\
        MOD & 9.80 & 6 & 0.251 & 0.66 & 2.13 \\
        \midrule
        \textbf{Overall} & \textbf{12.10} & \textbf{6} & \textbf{0.013} & \textbf{0.25} & \textbf{3.29} \\
        \bottomrule
    \end{tabular}
    \caption{Network statistics per community and overall. In order, the column names correspond to: Name of the community, average degree, diameter, density, average clustering coefficient, and average path length.}
    \label{tab:network_stats}
    \vspace{-0.5cm}
\end{table}

\begin{figure*}[t!]
    \centering
    \includegraphics[width=0.9\textwidth]{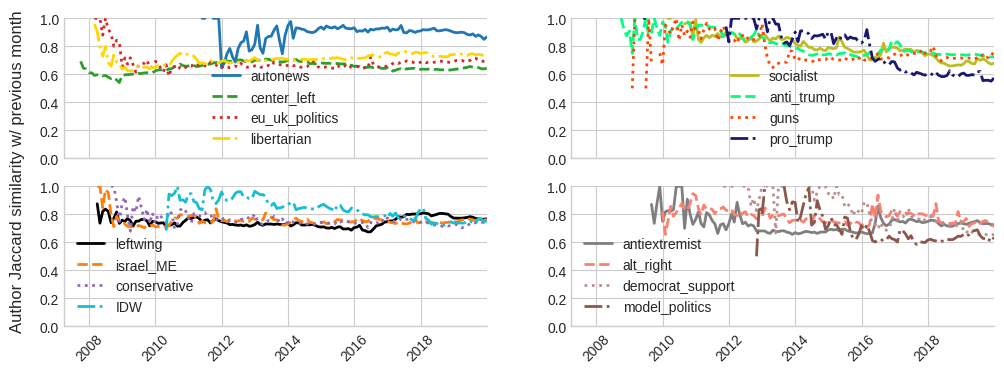}
    \caption{Time-series of Jaccard similarity between participating authors of any given month and the previous month.}
    \label{fig:jaccard_ts}
    \vspace{-0.2cm}
\end{figure*}

To ensure that individual subreddit bans did not substantially affect the aggregated communities, we conduct a time-series analysis where we obtain the Jaccard similarity between the sets of authors who appeared in a given month and its previous month in that community (Figure~\ref{fig:jaccard_ts}).
If subreddit bans drove users out of the entire community, there should be sharp similarity drops following the bans.
However, we observe no such drops, let alone drops following any of the 24 subreddits' ban dates.
Instead, we find somewhat erratic patterns near the start of the communities' formation when the numbers of participating authors were small, followed by convergence toward consistent similarities as the communities grew.
The sharp drop in the pro-Trump community around the start of 2016 is also attributable to a sudden growth in authors and activity that we observed in a separate analysis.
Generally, similarity values for every community, including those with and without banned subreddits alike, remained relatively high (above 0.6). 
This shows that community participants continued to be active in other subreddits in that community following the bans, and bans did not have substantial effects.

\subsection{User Commenting Prevalence}\label{sec:prevalence}

Next, we compute users' commenting prevalence, across the 16 communities, as the proportion of comments they have posted to that community.
We note that these prevalence values may be sensitive to cases where community moderators removed a substantial proportion of a user's comments.
For example, suppose a user who posted 100 comments had 55 comments in center-left community subreddits and 45 in pro-Democrat community subreddits. In that case, they get a score of 0.55 for CL, 0.45 for DEM, and 0 for the remaining 14 communities.

We assume that most users will engage predominantly with the communities they are part of, consistent with findings around homophily on social media~\cite{garimella_political_2018,zollo_debunking_2017}.
Thus, we derive each user's ``home'' community %
by taking the largest out of the 16 prevalence values (the majority community) for that user.

To filter out ``troll'' users who post spam or frequent communities with malicious intent (e.g., to harass or provoke), we only consider a user resident if their net upvotes are highest within that community {\em and} are above 1 (the default score of a newly posted comment).
This approach follows~\citet{an_political_2019,rajadesingan_political_2021}.

\subsection{Cross-community posting prevalence}

We plot a heatmap to show the average commenting prevalence of each community's home users in Figure~\ref{fig:heatmap}.

\begin{figure*}[t!]
    \centering
    \includegraphics[width=0.9\textwidth]{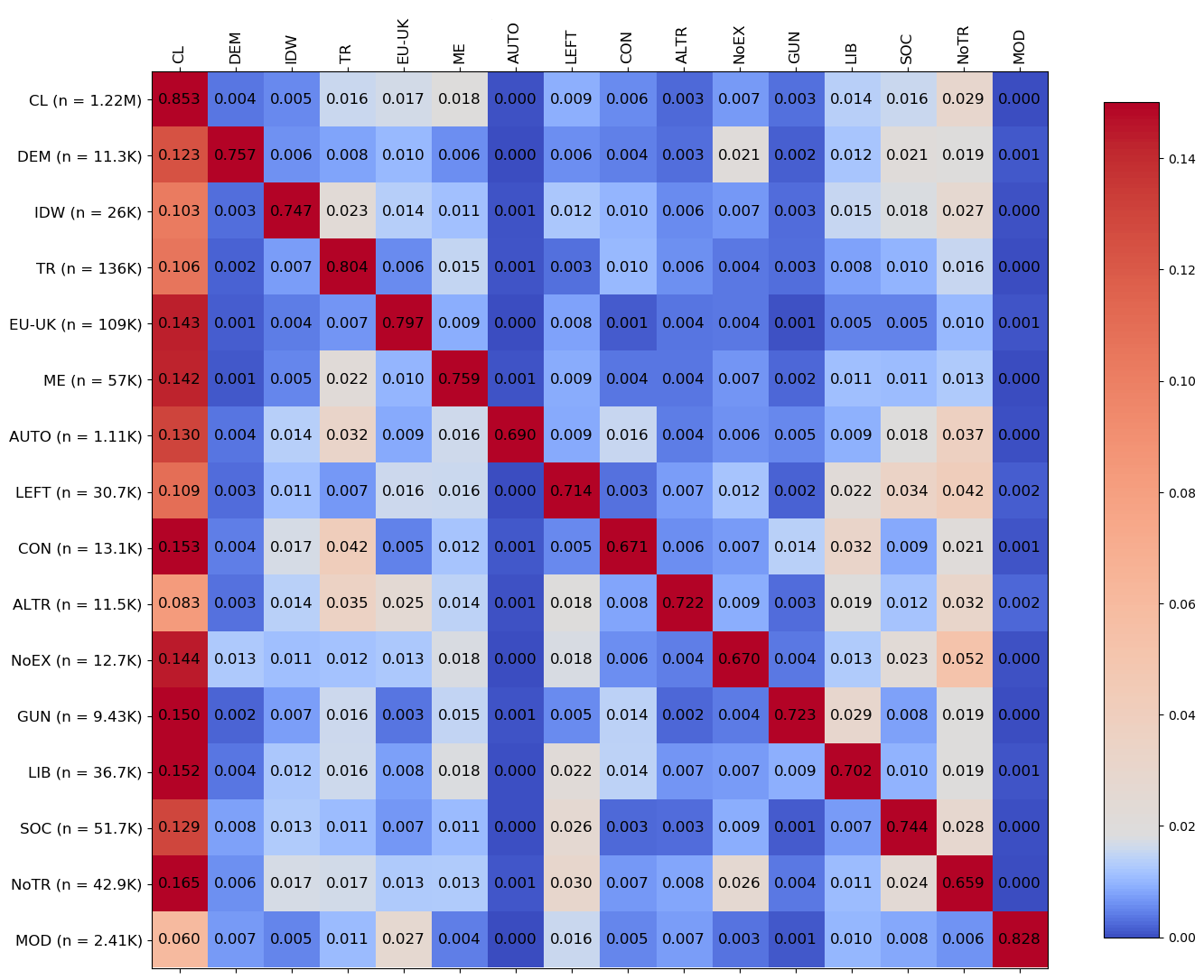}
    \caption{Heatmap of posting prevalence of each community's home users. n is the number of users with the respective community as their home {\em after} filtering. Reading across the horizontal shows outgoing posting prevalence to, and down the vertical shows incoming posting prevalence from, the respective communities' home users.}
    \label{fig:heatmap}
\end{figure*}

Values represent the average percentage of comments home users in the row community leave on the respective column community.
Center-left, which is by far the largest and most mainstream community, draws a fair degree of traffic from virtually all other communities.
There is also moderate traffic between some ideologically aligned communities (e.g., Conservative and alt-right to pro-Trump, Socialist and anti-Trump to left-wing).
In some cases, we also observe mild-to-moderate traffic between adversarial communities (e.g., alt-right to anti-Trump).
Overall, users have pretty diverse commenting prevalences outside of, but still predominantly engage with, their home communities.

\section{Toxicity Analysis}\label{sec:main_analysis}

This section analyzes how commenting prevalence is related to toxic behavior on Reddit.
We focus on how users' involvement in their home community influences how toxic they are elsewhere.
Also, we examine whether it is a community's home or non-home users who drive toxic discussions and shed light on toxicity relationships between communities.

\descr{Toxicity.} 
We use Perspective API's Severe Toxicity model to label comments as toxic or non-toxic.
Severe Toxicity is defined as ``a very hateful, aggressive, disrespectful comment or otherwise very likely to make a user leave a discussion or give up on sharing their perspective.''\footnote{\href{https://support.perspectiveapi.com/s/about-the-api-attributes-and-languages}{https://support.perspectiveapi.com/s/about-the-api-attributes-and-languages}}
This provides a score between 0 and 1, and we consider a comment to be toxic if its Severe Toxicity score is above 0.7.\footnote{\href{https://support.perspectiveapi.com/s/about-the-api-score}{https://support.perspectiveapi.com/s/about-the-api-score}}

Although not free from important limitations, e.g., sensitivity to adversarial text~\cite{jain_adversarial_2018,hosseini_deceiving_2017} and bias toward text mentioning marginalized groups or written in African-American English~\cite{sap_risk_2019}, Perspective outperforms alternative models~\cite{zannettou_measuring_2020} and allows us to measure relative toxicity at scale.

\descr{Overview of toxicity patterns.}
As our dataset spans a wide range of time ($>$ 13 years), we provide toxicity time-series plots to capture a) how toxicity behaviors evolve in each community and b) when each community begins to materialize on Reddit (Figure~\ref{fig:all_timeseries}).

\begin{figure*}[!t]
    \centering
    \includegraphics[width=0.85\textwidth]{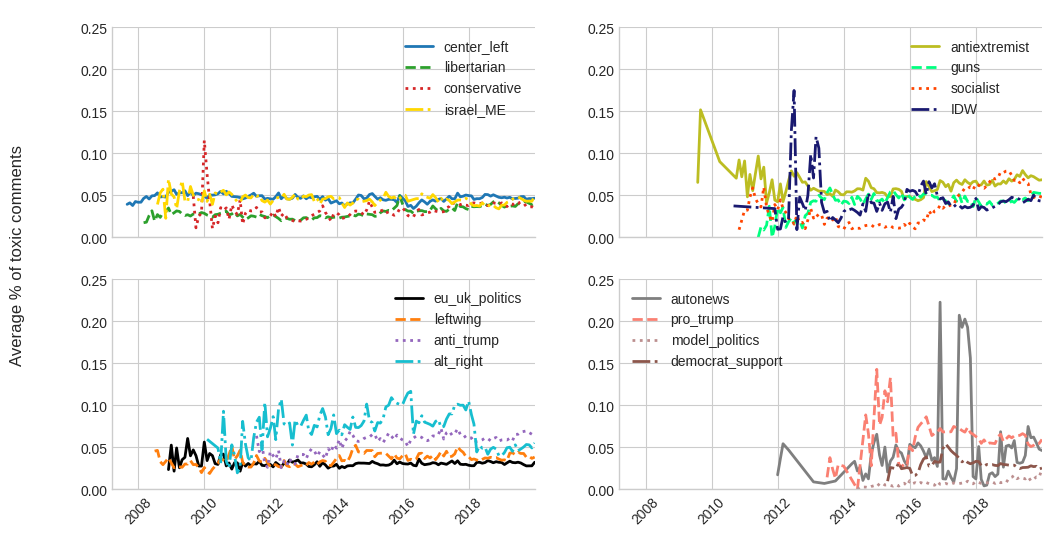}
    \caption{Time-series of the proportion of toxic comments in each community at a monthly granularity (e.g., 0.25 = 25\% of comments in that month are toxic). Months with fewer than 100 total comments in the respective community are not shown.}
    \label{fig:all_timeseries}
\end{figure*}

We observe that some communities are rooted in Reddit even before finding their ``purpose''.
For example, the anti-Trump and pro-Trump communities appear in mid-2011 and mid-2013, respectively, long before Donald Trump's presidential campaign was officially launched on June 16th, 2015.
This potentially indicates the presence of subreddits, which then aligned with different perspectives on more contemporary political events.

Some communities are relatively stable in the toxicity they display over time (e.g., center-left and left-wing).
Others, e.g., IDW and pro-Trump, show toxicity spikes at specific points in time but re-stabilize.
These are possibly event-driven, e.g., the announcement of Donald Trump's presidential run in mid-2015 for the pro-Trump community.

\subsection{Predictors of Community Toxicity}

First, we examine the overall toxicity of both home and non-home users in focal communities and the toxicity of home users in other communities.
For every user, we calculate the proportion of toxic comments on each community they have posted.
To preserve variability, we exclude users with fewer than ten total comments or fewer than five comments on the respective target community.
We plot this in Figure~\ref{fig:tox_bars}.

Model politics (MOD) was the least toxic across all three groups.
Alt-right (ALTR) drew the most toxicity from home and non-home users.
Anti-Trump (NoTR) was the most toxic in other communities.
In all communities, except center-left (CL), pro-Trump (TR), and socialist (SOC), non-home users were at least as toxic, if not more, as home users.

The green bars in Figure~\ref{fig:tox_bars}, which represent the toxicity of home users in other communities, show that the same users may change their toxic behavior depending on which community they are posting in.
For all but three communities (DEM, Intellectual Dark Web (IDW), GUN), these bars are either higher than or lower than {\em both} home and non-home users' toxicities (i.e., they are not higher than one and lower than the other).
This means that in ALTR, anti-extremist (NoEX), and auto-news (AUTO), the most toxic communities overall, users modified their behavior when posting elsewhere.
This could be due to better moderation elsewhere, user self-regulation, or, more likely, a combination of both.
On the contrary, left-wing (LEFT) and SOC home users became more toxic when posting elsewhere despite these communities being low on toxicity.
Our results suggest that community norms influence toxicity levels beyond individual users' tendencies, consistent with the findings of~\citet{rajadesingan_quick_2020}.

\begin{figure}[t]
    \centering
    \includegraphics[width=0.99\columnwidth]{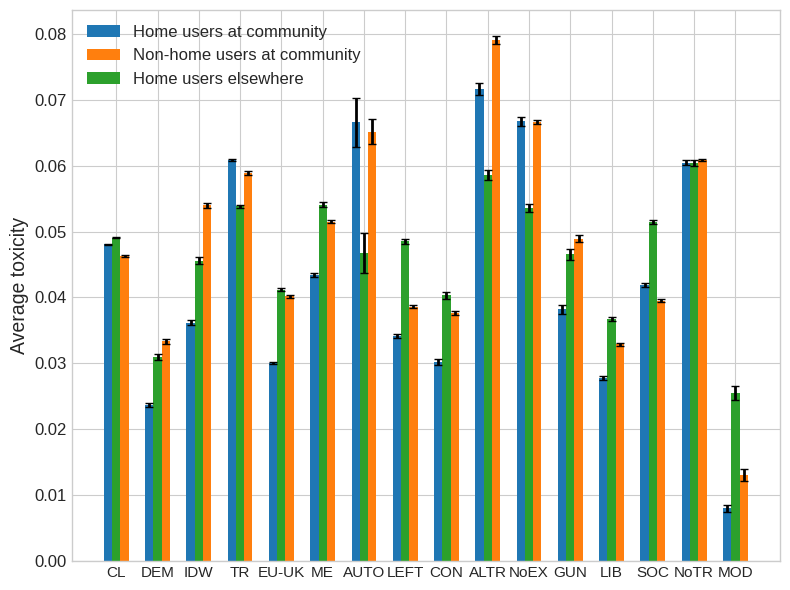}
    \caption{Average toxicity of home and non-home users per community and average toxicity of home users on other communities. Error bars represent standard errors.}
    \label{fig:tox_bars}
      \vspace{-0.2cm}
\end{figure}

\subsection{Pairwise Community Toxicity}\label{subsec:pair_tox}

We examine pairwise toxicity relationships between communities.
Rather than taking the average toxicity of each home user, we now pool all comments posted from a community's home users on another one and compute the proportion of toxic comments out of the total comments in the pool.
Figure~\ref{fig:tox_heatmap} is a pairwise toxicity proportion heatmap.

\begin{figure*}[t!]
    \centering
    \includegraphics[width=0.9\textwidth]{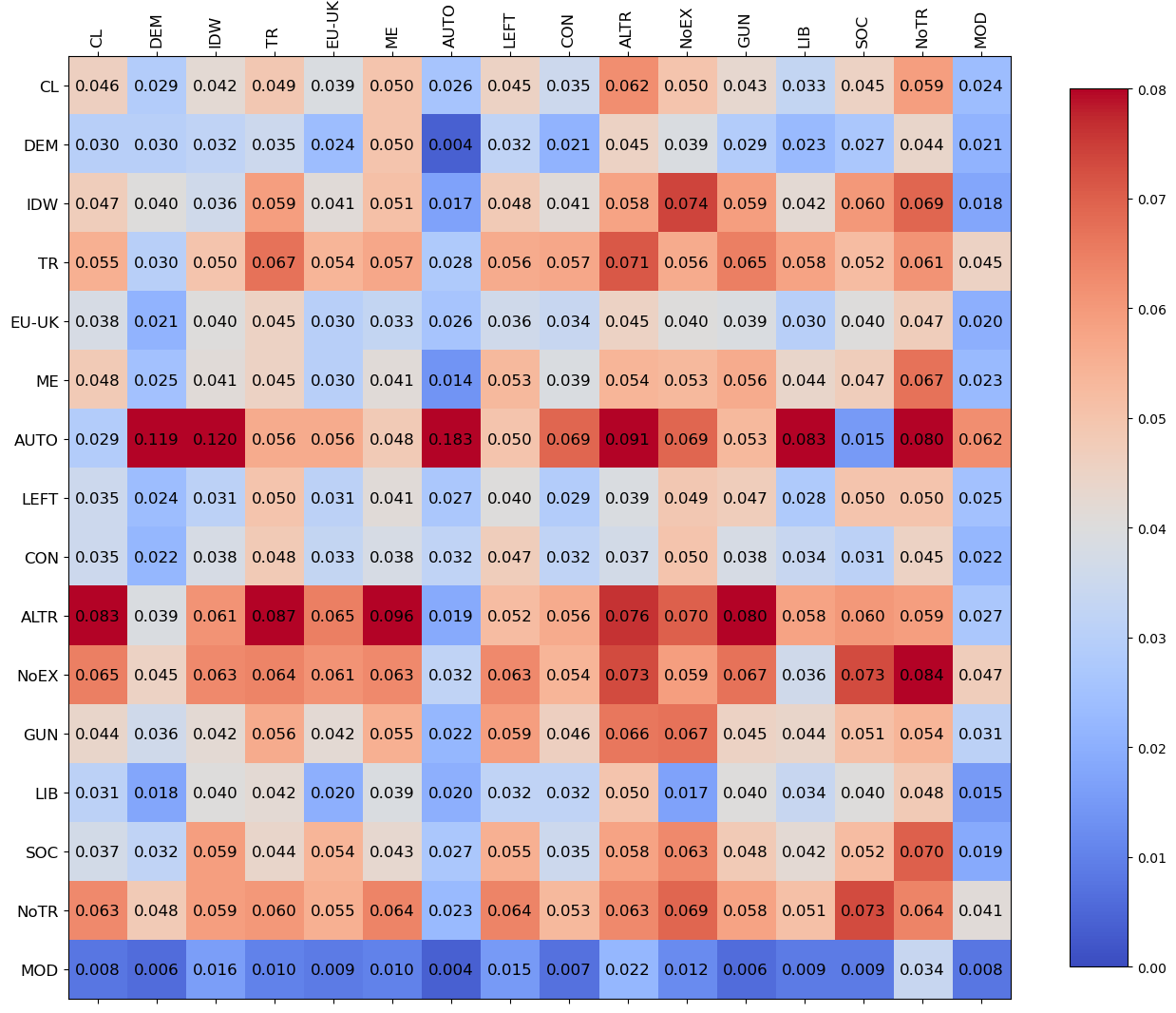}
    \caption{Heatmap showing the pairwise proportion of toxic comments posted by the communities' home users. Outgoing toxicity is shown across the horizontal, and incoming toxicity is shown down the vertical.}
    \label{fig:tox_heatmap}
        \vspace{-0.3cm}
\end{figure*}

The heatmap follows the user-level toxicity patterns in Figure~\ref{fig:tox_bars}.
For example, ALTR and NoEX show much higher toxicity with most comparisons, both in terms of outgoing and incoming toxicity.
Similarly, pro-Democrat (DEM) and MOD show lower toxicity across the board.

However, Figure~\ref{fig:tox_heatmap} also demonstrates that toxic behavior is not inherently {\em tribal.}
That is, we observe high toxicity between communities on the same side of the political spectrum, e.g., ALTR to pro-Trump (TR) and Socialist (SOC) to NoTR.
Similarly, some communities on opposing sides of the political spectrum show lower cross-toxicity---e.g., DEM and Conservative (CON) (in both directions).
Nonetheless, moderation potentially plays a vital role in these patterns (e.g., selective moderation of toxic comments based on the commenter's political leaning).

\subsection{Association of Echo Chamber Engagement with Non-Home Toxicity}

Next, we quantify the relationship between echo chamber prevalence and toxicity displayed in other communities.
Our goal is to assess, for each possible community pair, whether the posting prevalence of users in the home community was related to the toxicity of their comments at the target. 

\descr{Modeling.}
We treat each comment as a single observation. 
Every comment posted by a user {\em outside their home} is a Bernoulli trial, and a ``success'' is a {\em toxic} comment.
We then set each user's home community {\em a posteriori} as described in Section~\ref{sec:prevalence}.
Thus, our model is limited in assuming that users do not change their homes over time.
We observe each user's comment trail and dynamically update their home posting prevalence based on how many comments they have posted at home and non-home up to that point.

We treat individual user IDs as nesting variables, maintaining independent observations between users and dependence between comments from the same user.
We then run mixed-effects logistic regressions for each pairwise community comparison, allowing the slope and intercept of each user to vary as random effects~\cite{bates_fitting_2015}:
\begin{equation*}
\small
\begin{split}
P(toxicity_{c\in T}) = logit&( \beta_0 + \beta_1 prevalence_H + b_{0i} + \\
                                & + b_{1i} prevalence_H + \varepsilon)
\end{split}
\end{equation*}
where T is the set of comments in the target community posted from the home community's users in the pairwise comparison, and H is the home community.
The model is a mixed effect one because we take repeated measures from each user every time they post in a non-home community, and we derive the fixed probability that a comment will be toxic based on echo chamber (home posting) prevalence and whether the comment is toxic at the time of each observation.

\descr{Results.} We plot all pairwise regression estimates ($\beta_1$ log odds values) in a ``faux'' forest plot (Figure~\ref{fig:forest}).
On average, each comparison consists of 327K comments from 19.5K individual authors.
The smallest comparison is 2.01K comments from 308 authors (GUN to DEM), while the largest is 5.97M comments from 390K authors (CL to NoTR).
In comparisons where the model fails to converge, we test three different optimization algorithms (nlminb~\cite{gay_usage_1990}, L-BFGS-B~\cite{zhu_algorithm_1997}, Nelder-Mead~\cite{nelder_simplex_1965}), and report confidence estimates with successful convergence.
Convergence failures are marked with ``cf.''

\begin{figure*}[t]
    \centering
    \includegraphics[width=0.99\textwidth]{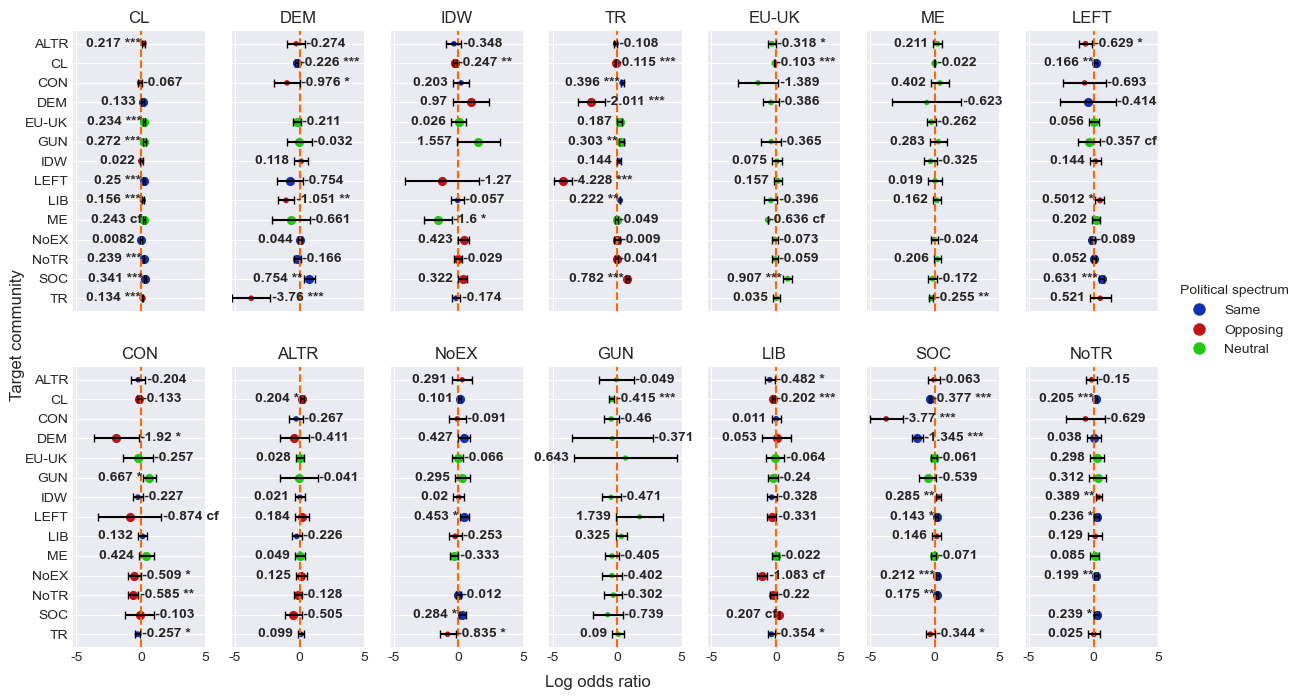}
    \caption{Forest-like plot showing pairwise regression estimates of echo chamber prevalence on toxicity probability at target community. *\textit{p} $<$ 0.05, **\textit{p} $<$ 0.05/13, ***\textit{p} $<$ 0.05/182. cf = convergence failure. Error bars represent 95\% confidence intervals. Community relation is the same if both communities are right-leaning (IDW, TR, CON, ALTR, LIB) or both are left-leaning (CL, DEM, LEFT, NoEX, SOC, NoTR), opposing if one community is left-leaning and the other right-leaning, and neutral otherwise (at least one neutral community). Dot size is based on the number of observations relative to the largest number of observations. Notice the log-odds scale.}
    \label{fig:forest}
        \vspace{-0.2cm}
\end{figure*}

We omit pairwise comparisons where the home and target community is the same due to the high number of users (individual slopes) in these data, which introduces computational restraints.
Furthermore, we omit the AUTO and MOD communities from these analyses because their low numbers of home users do not provide adequate statistical power.

We report significance at three different levels: 1) an $\alpha = 0.05$ cutoff, 2) a Bonferroni-corrected $\alpha = 0.05/13$ cutoff for multiple comparisons using the same population (since each population of home users is used 13 times in comparisons), and c) a hyper-conservative $\alpha = 0.05/182$ cutoff point for all comparisons in the plot.
We use level a) for our interpretations in the remainder of this paper as we are interested in the unique relationships between each pair, although level b) may also be reasonable.
Level c) is only used for transparency purposes; we do not recommend it as it is over-corrective and can inflate Type II errors.
 
\descr{Discussion.}
Positive values show ``polarizing'' associations (i.e., higher home posting associated with a higher probability of toxicity at target), while negative values show ``depolarizing'' associations.

There are no universal associations based on whether the communities are on the same or opposing sides of the political spectrum; associations are unique to each pair.
Furthermore, relationships are not necessarily reciprocal.
For example, SOC users become less toxic in the DEM community as they post more at home, while the opposite holds for DEM users on SOC.
In an oppositional pair example, increased echo chamber prevalence in TR makes toxicity more likely on SOC, while the opposite is true the other way around.

The largest polarizing relationship (EU-UK to SOC) amounts to a user posting close to 100\% at home being about 2.5 times as likely to be toxic at the target compared to someone having posted nothing at home.
Toxicity likelihood is drastically less likely (0.015 times) in the largest depolarizing relationship (TR to LEFT).
This may again be attributable to TR users being heavily moderated on LEFT.
Overall, the directions and effect sizes vary heavily for unique community pairs.
However, toxicity \emph{associations} with increased engagement in the home community remain separate from the \emph{actual} toxicity displayed by home users in another community.
That is, we may observe flat effects on communities that are otherwise toxic to each other.
This is a distinction we clarify in the next section.

\section{Typology of Community\\Relationships}\label{sec:typology}
Thus far, our analyses have focused on how political Reddit communities are related to one another in terms of their toxicity and echo chamber prevalence.
Now, to better understand these cross-community dynamics, we synthesize our findings into a coherent typology based on three dimensions:
\begin{enumerate}
\item {\em Cross-community toxicity} (Figure~\ref{fig:tox_heatmap}).
We define the relationship as:
    \begin{compactitem} 
    \item inciting, if cross-toxicity $\geq0.056$ (highest quartile),
    \item composed, if $\leq0.031$ (lowest quartile), and 
    \item basic, if $0.031 < toxicity < 0.056$.
    \end{compactitem}
\item {\em Increased engagement in the home community} (Figure~\ref{fig:forest}). 
We define the relationship,    with significance interpreted at $\alpha = 0.05$,  as: 
    \begin{compactitem} 
    \item polarizing, if the pair model is significant and positive,
    \item  depolarizing if the model is significant and negative, %
    \item non-effectual if the model is non-significant (or cf, convergence failed). 
    \end{compactitem}
\item {\em Agreement in political leaning.} 
This is done based on a qualitative assessment of the communities' constituent subreddits, as discussed in Section~\ref{sec:communities}. 
We define this as:
    \begin{compactitem} 
    \item same, if both communities are right- or left-leaning, 
    \item opposing if one community is right- and the other left-leaning, and 
    \item neutral otherwise.
    \end{compactitem} 
\end{enumerate}

\begin{figure*}[t]
    \centering
    \includegraphics[width=0.93\textwidth]{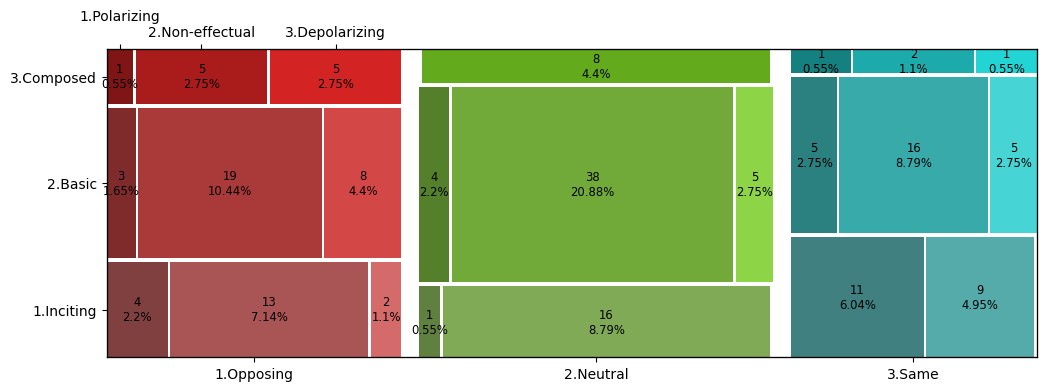}
    \caption{Mosaic plot showing number and proportions of typologized community relationships. Left axis: Type of speech based on cross-toxicity. Bottom axis: Political spectrum leaning. Top axis: Association with increased engagement at home.}
    \label{fig:mosaic}
        \vspace{-0.2cm}
\end{figure*}

\subsection{Typology frequencies}
Figure~\ref{fig:mosaic} is a mosaic plot showing the frequency of community relationship types.
The most common type was an indifferent one (basic, non-effectual, and neutral) at 20.88\% of the pairwise comparisons.
By proportion, the basic and non-effectual types were more common among neutral pairs than opposing or same-spectrum pairs. 
However, basic and non-effectual was still the most common type within the subsets of opposing (e.g., LEFT to TR; LIB to SOC; ALTR to LEFT) and same-side (e.g., NoTR and DEM both ways; CON and LIB both ways) pairs.

Interestingly, inciting and polarizing types were more common in same-spectrum community pairs.
Out of the 11 such same-side relationships observed, nine occurred on the left, with the main ``perpetrator'' communities being SOC (to NoTR and NoEX), NoEX (to LEFT and SOC), and NoTR (to CL, SOC, LEFT, and NoEX). 
CL to NoTR was the remaining relationship of this type on the left.
DEM was the only community that was neither an originator nor a receiver of this type on the left.
The TR community was the sole originator of this type on the right (to LIB and CON).
However, we also observe inciting and polarizing relationships with opposing-side pairs on four occasions (CL and ALTR both ways; SOC to IDW; NoTR to IDW).
Additionally, depolarizing and composed relationships were most common among opposing pairs (e.g., DEM and CON both ways; Libertarian to CL) rather than same-side pairs; the only depolarizing and composed same-side pair was DEM to CL.
This suggests that echo chamber-driven animosity may have predominantly occurred in politically agreeable communities.

Users were more likely to demonstrate hostility toward political outgroups when they interacted with ideologically aligned others (or within ideologically aligned communities) rather than when directly interacting with counter-partisans, which somewhat clashes with the conventional wisdom that toxic behavior is more common under political misalignment.
To a lesser extent, ideologically aligned individuals also directed hostility toward each other due to ``in-fighting'' (see Section~\ref{sec:annotation}).
However, as mentioned in Section~\ref{subsec:pair_tox}, we stress that many of these patterns may be what \textit{remained} on the communities following moderation, which leaves the possibility that toxic comments were selectively moderated based on the commenters' leaning.

We also observe some rare ``wild card'' types.
Simultaneously inciting and depolarizing relationships (which only occurred with the opposing pairs TR to LEFT and NoEX to TR) suggest that there may exist learned civility amidst otherwise inciting discourse or that more frequent origin-community users may be more likely to have their comments removed on the target communities.
Similarly, composed but polarizing relationships (DEM to SOC for same-side; LEFT to LIB for opposing) show that increased activity may lead to higher toxicity, even in otherwise civil discourse.

Overall, we find just one case of a depolarizing and composed relationship on the same side of the political spectrum, while we observe five such relationships for opposing-side pairs.
At the same time, same-side pairs were more likely to have inciting and polarizing relationships, especially among left-wing communities (except for DEM, which was primarily involved in composed and depolarizing relationships).
However, we also note that an inciting {\em and} polarizing relationship may not necessarily be more problematic than, say, {\em just} an inciting one.
For example, ALTR, which was one of the most extreme communities in our dataset, was the originator of 7 inciting but non-effectual types; this means that ALTR users tended to be more toxic on many other communities, but they did not become {\em even more} toxic as they posted more at home.
This could be due to, e.g., the ALTR community already being very high in toxicity, which would leave a smaller margin for an increase in toxicity levels.

\subsection{Annotation study}\label{sec:annotation}
Next, we perform an annotation study to clarify whether the cross-toxicity among same-leaning communities points to political in-fighting between these communities or ``ganging up'' to reprimand the political outgroup collectively.

First, we create two data pools of toxic (as per Perspective API) left-to-left and right-to-right comments.
Every comment in these pools is a top-level comment, i.e., a direct response to the submission, because deeper-level comments tend to lack the crucial context required to understand their target.

We then randomly sample 400 comments, 200 from each pool.
We extract the title, body (if any), hyperlinks (if any), and media such as images or videos (if any) included in the submission each comment is responding to for further context around the comment.
Each comment is labeled by two annotators (two authors of this paper) as per two categories: toxicity (toxic or non-toxic) and target (outgroup-directed, ingroup-directed, unclear).
For toxicity, we obtain good agreement (Krippendorff's $\alpha$ of 0.71 and 0.83 for left and right, respectively).
The vast majority of comments in the pools are confirmed by the annotators to be toxic (87.5\% and 85.5\% for left and right, respectively).
For the target, we obtain moderate $\alpha$ ratings of 0.59 and 0.52 for left and right, respectively.
We resolve disagreements through a discussion of contested comments.

In Figure~\ref{fig:annotation_targets}, we report the results of the annotation. 
(Note that we omit misclassified non-toxic comments as they do not fall in the study's aims).
Ignoring comments with unclear targets, we find that the vast majority of toxic comments are indeed directed toward political outgroups, suggesting that the polarizing patterns we observe may be primarily due to same-leaning communities instigating rather than attacking each other.
However, there was also a non-negligible proportion of comments (25.2\% and 16.7\% for left and right, respectively) demonstrating political in-fighting.
This primarily reflects disagreements on endorsed politicians, issue positions, or clashes between ideologies (e.g., anarchism vs. state socialism).

\begin{figure}[t]
    \centering
    \includegraphics[width=0.85\columnwidth]{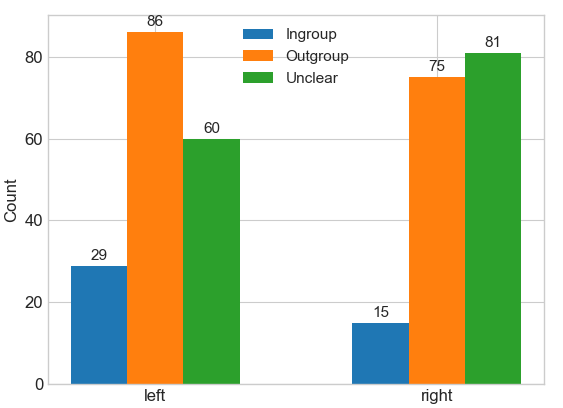}
    \caption{Targets of same-to-same leaning toxic comments as determined through annotations.}
    \label{fig:annotation_targets}
        \vspace{-0.4cm}
\end{figure}

Consistent with previous patterns, in-fighting is somewhat more frequent among the left.
Although the annotation study concerns a much smaller scale than our previous analyses, it confirms that polarization may occur mostly when same-leaning users interact with each other and speak negatively about political outgroups in those outgroups' absence.

\section{Discussion \& Conclusion}

This work presents a large-scale, historical analysis of Reddit's political spaces between 2006 and 2019.
We aim to determine whether the degree of engagement with echo chambers relates to behavior outside them.
We find that political communities on Reddit were more varied than the traditional left-right split during this period.
Each community carried its norms in the toxicity of conversations it hosted and how its users behaved elsewhere.

Users predominantly engaged with their home communities, consistent with echo chamber perspectives, but they also posted a non-negligible number of comments to other communities.
For {\bf RQ1}, which concerns how echo chamber engagement relates to the probability of hostile intergroup interactions, we find whether the degree of echo chamber (i.e., home) engagement related to toxicity in a target community depended on the unique relationship between the two communities.
That is, increased or decreased polarization could occur between and within political leanings, with these relationships not necessarily reciprocal.

For {\bf RQ2}, regarding how communities' polarization and hostility relationships vary based on political leaning, typologizing the communities revealed interesting patterns.
Surprisingly, inciting and polarizing types were more common between communities on the same side of the political spectrum; however, this mainly reflected the reprimand of political outgroups rather than in-fighting.
The presence of ``wild card'' combinations (e.g., polarizing and composed, inciting and depolarizing) suggests that political discourse is complex and influenced both by established cross-community and individual users' engagement patterns.
Different communities had unique relationships, and echo chamber engagement did not act in unilateral directions.
Nonetheless, content moderation possibly played a substantial (albeit unclear) role in the patterns observed.

\subsection{Implications}

Our work is a first attempt at bridging the echo chamber and hostile interaction perspectives of polarization, exploring how the two may be interdependent. 
Furthermore, it is among the first studies looking at the \textit{degree} of echo chamber engagement at the user level rather than focusing on distinct chamber-like communities. 
The complex picture from our study suggests that increased engagement with specific communities can broadly be associated with both polarization and depolarization of users.
We also found more cross-polarization among left-wing communities, and this was mostly outgroup- (i.e., right-wing-) directed.

At the same time, we also observed a higher degree of in-fighting among left-wing communities, which might partly explain why left-wing radicalization is less common than right-wing radicalization~\cite{hosseinmardi_evaluating_2020,ribeiro_auditing_2020} as left-wing users encounter more attitudinal disruption, and this may keep more extreme opinions in check.
Arguably, our findings are important for several reasons.

\descr{High-level mapping.}
First, they situate user activity in the broader context of Reddit's political sphere.
Our dataset captures a relatively complete set of {\em political} subreddits, which allows us to get a broader picture than possible by studying select subreddits over more restricted periods.
Thus, we can capture a high-level overview of the complex interplay between engagement patterns and toxic behavior and highlight polarizing cases where diversifying user engagement could reduce hostile interactions.
A potentially fruitful direction to mitigate polarization could be to focus on organically occurring communities that ostensibly increase the diversity of engagement (e.g., {\em r/changemyview} or {\em r/AskTrumpSupporters}) and design systems that encourage and support more communities to form.

\descr{Indifferent communities.}
Second, we show that several communities within Reddit's political space were fairly neutral and indifferent toward each other in all respects; indeed, this was the most common type of relationship.
Given that such communities hosted political discussions which were not particularly charged, they may be studied for their potential as ``online buffer zones'' where users' stances on individual issues, rather than their political leanings, are most salient.
Overall, this pattern suggests that political polarization is a more contextual rather than the ubiquitous problem.
However, especially regarding cross-toxicity, our typology was {\em relative} to other relationships (i.e., only 25\% of relationships could be treated as inciting due to quartile-based classification).
At the same time, other work has found that political discussions, in general, tend to be more toxic than other kinds of conversations~\cite{rajadesingan_political_2021}; therefore, even our lower-toxicity relationships could be more toxic than relationships in domains other than politics.
Moreover, content moderation may have distorted our picture of real-time polarization.
Further work is needed to verify how contextual polarization truly is.

\descr{Polarization in aligned communities.}
Third, while past work~\cite{cinelli_echo_2021,marchal_be_2021,garimella_political_2018} has focused on echo chambers and hostile interactions between counter-partisans as explanations for polarization, here we find inciting and polarizing patterns predominantly between politically aligned communities (especially among the left), but composed and depolarizing patterns predominantly between politically opposed ones.
One potential explanation is that hostile interactions between opposing partisans may be more context-specific than previously thought, as they did not dominate when examining the broader community context.
Another explanation is that toxic behavior may occur in real-time; however, this is retroactively moderated selectively only when this behavior comes from users whose views disagree with the broader community.
Indeed, the likelihood of comment removals increases drastically when a community is negatively targeted by another on Reddit~\cite{kumar_community_2018}.
Regardless, polarization in our dataset was observed largely in the form of agreeable communities inciting and reinforcing each other when speaking negatively about political outgroups.

These results could also carry important implications for content moderation.
Mainly due to ideological biases or subreddit-specific rules, users aligning with a community's political stance may be allowed to continue displaying toxic behavior as long as they do not cross partisan lines. 
In turn, this can result in evocative polarization even without ideological opponents.
This is an important consideration that warrants future work, as it raises potential questions around the differences between stated and realized moderation goals (e.g., whether it is anti-hostility or anti-dissent).

\subsection{Limitations and Future Work}

\descr{Engagement vs. exposure.}
Although we start from the idea of exposure to diverse information, what we measure is commenting activity (i.e., engagement).
Given that we derive our communities using author similarity, we approximate exposure by assuming that similar subreddits will host similar opinions as they feature similar users.
However, many users may ``lurk'' in oppositional political spaces and view but never engage with posts.
Therefore, we hope that future work will study the polarization phenomenon in the context of true information exposure, using metrics like clicks and reading time of different pieces of content (see, for example, \cite{garimella_political_2021}).

\descr{Hostility as toxicity.}
Our measure of toxicity may arguably only represent a small part of possible expressions of hostility; others include, for example, anger~\cite{kumar_community_2018}, inter-community attacks~\cite{kumar_community_2018,datta_extracting_2019}, or negative sentiment~\cite{de_francisci_morales_no_2021}.
Furthermore, given our findings from the annotation study, this ``hostility'' may not necessarily be hostility toward the target community per se but rather a third outgroup community altogether.
Future research could examine several such measures of hostility alongside each other (e.g., anger, toxicity, etc.) when studying inter-community relationships to observe which expressions are the most dominant.

\descr{Non-causal inference.}
Our model uses time data to observe an effect (toxicity) following a previous event (home-community posting prevalence); however, this was not a true causal effect since other factors could be driving both toxicity and posting prevalence (e.g., the interest of the user in less controversial topics, the radicalization of the user on other platforms, etc.)
Therefore, future research could employ methods more suited to causal inference, such as controlled experiments or regression discontinuity analysis.

\descr{Content moderation.}
We aimed to study cross-toxicity between communities and whether this toxicity was more pronounced for users who demonstrated more one-sided engagement with their preferred communities.
However, we did not clarify whether these patterns were due to moderation measures or naturally occurring.
Polarized communities may have been more toxic due to more lax moderation, which could bias our results.
Future research could distinguish between these two scenarios, as this is important for understanding how interactions of any type (i.e., intergroup or intragroup ones) arise online for other users to witness.

\descr{Selection of subreddits.}
We intentionally chose a wide range of subreddits to cluster based on the amount of political content they host to match the large scope of our research questions.
However, in doing so, we also lost some qualitative information regarding these spaces.
For example,~\citet{an_political_2019} only studied four subreddits, but these were carefully selected based on the specific political candidates they supported, whether contrarian discourse was allowed on the subreddit, and other unique characteristics.

Some of the specific inter-community relationships we observed might be due to the unique characteristics of these communities; for example, some may have predominantly hosted subreddits that advocated for specific political candidates, and others may have been pro- or anti-establishment, etc.
Furthermore, the period of our observation period was very large (13 years).
Throughout this period, some users could have changed their political affiliations or issue positions, and some subreddits were banned (although we did not find any substantial effects of these bans on the aggregated communities).
The long time span also opens the possibility that various events could have taken place that affected activity on Reddit and any communities which were active at the time but were not considered here.

Further, some of the subreddits may have been clustered together based on their ideological agreement, whereas others may have been clustered together simply based on the topic (even though they may disagree on several issues).
With US-based spaces, we mainly observed the former (except for the Gun community, which hosted both pro- and anti-gun subreddits).
However, as with the European community, we saw both right-leaning and left-leaning subreddits clustered together.
In turn, interactions with other communities may have occurred only between subreddits that were ideologically aligned~\cite{soliman_characterization_2019}.

While the scale of our analysis was a methodological choice to generalize our findings beyond specific cases, future research could adopt a more qualitative period and community selection method to determine when and for which communities the different types of relationships hold.
This is particularly important considering that the treatment of ``echo chambers'' in this study was relatively broad, and discourse within these bundled subreddits was likely more diverse than what would typically be expected in traditional echo chambers.
Indeed, the different communities we studied possibly hosted different levels of ideological homogeneity, and it is likely that resident users, especially among the larger communities, held different viewpoints on several issues. %
We hope future work can probe this within more ideologically homogeneous spaces.

\subsection*{Acknowledgments}
We thank Ashwin Rajadesingan for the initial list of 31K subreddits with political prominence labels.
We also thank the anonymous reviewers for their valuable feedback.

This work was partially funded by the UK EPSRC grant EP/S022503/1, which supports the UCL Centre for Doctoral Training in Cybersecurity, the UK's National Research Centre on Privacy, Harm Reduction, and Adversarial Influence Online (REPHRAIN, UKRI grant: EP/V011189/1), and the US NSF under grants IIS-2046590, CNS-2114411, CNS-1942610, and CNS-2114407.
Any opinions, findings, conclusions, or recommendations expressed in this work are those of the authors and do not necessarily reflect the views of the funders.

\bibliographystyle{abbrvnat}

\end{document}